\documentclass[aps,pre,twocolumn,showpacs]{revtex4}
\usepackage{epsfig,graphicx}
\draft
\begin{document}

\title{Does the scale-free topology favor the emergence of cooperation?}

\author{Zhi-Xi Wu \footnote {Electronic address: wupiao2004@yahoo.com.cn}, Xin-Jian Xu, and Ying-Hai Wang \footnote {Electronic address: yhwang@lzu.edu.cn}}

\address{Institute of Theoretical Physics, Lanzhou University, Lanzhou Gansu 730000, China}

\date\today

\begin{abstract}
In a recent Letter [F.C. Santos and J. M. Pacheco Phys. Rev. Lett. \textbf{95},
098104 (2005)], the scale-free networks are found to be advantageous for the
emergence of cooperation. In the present work an evolutionary prisoner's
dilemma game with players located on Barab\'asi-Albert scale-free networks is
studied in detail. The players are pure strategist and can follow two
strategies: either to defect or to cooperate. Serval alternative update rules
determining the evolution of each player's strategy are considered. Using
systematic Monte Carlo simulations we have calculated the average density of
cooperators as a function of the temptation to defect. It is shown that the
results obtained by numerical experiments depend strongly on the dynamics of
the game, which could lower the important of scale-free topology on the
persistence of the cooperation. Particularly, the system exhibits a phase
transition, from active state (coexistence of cooperators and defectors) to
absorbing state (only defectors surviving) when allowing \lq\lq worse\rq\rq
strategy to be imitated in the evolution of the game.
\end{abstract}
\maketitle

Cooperation is widespread in many biological, social and economic systems
\cite{Maynard}. Understanding the emergence and persistence of cooperation in
these system is one of the fundamental and central problems \cite{Maynard,
Dugatkin}. In investigating this problem the most popular framework utilized is
game theory together with its extensions to an evolutionary context
\cite{Hofbauer}. One simple game, the Prisoner's Dilemma game (PDG), has
attracted most attention in theoretical and experimental studies \cite{Gintis}.
In the standard PDG, the players can either defect or cooperate; two
interacting players are offered a certain payoff, the reward $R$, for mutual
cooperation and a lower payoff, the punishment $P$, for mutual defection. If
one player cooperates while the other defects, then the cooperator gets the
lowest sucker's payoff $S$, while the defector gains the highest payoff, the
temptation to defect $T$. Thus, we obtain $T>R>P>S$. It is easy to see that
defection is the better choice irrespective of the opponent's selection. For
this reason, defection is the only evolutionary stable strategy in fully mixed
populations \cite{Hofbauer}.

Since the cooperation is abundant and robust in nature, considerable efforts
have been expended trying to understanding the evolution of cooperation on the
basis of the PDG \cite{Dugatkin, Hofbauer, Gintis, Nowak, Nowak_1, Szabo_1,
Zimmermann, Tomochi, Santos}. These extensions include those in which the
players are assumed to have memory of the previous interactions \cite{Nowak},
or players are spatially distributed \cite{Nowak_1}, or allowing the players to
voluntary participating \cite{Szabo_1}. In addition, dynamic network model
\cite{Zimmermann} and dynamic payoff matrices \cite{Tomochi} were also
introduced to sustain high concentration of cooperation in PDG. In a recent
Letter \cite{Santos}, Santos and Pacheco have studied the PDG and another
famous game, the snowdrift game (commonly known as the hawk-dove or chicken
game), on scale-free networks and observed interesting evolutionary results:
due to the underlying network generated by growth and preferential attachment
(or the scale-free topology), the cooperation can be greatly enhanced and
becomes the dominating trait throughout the entire range of parameters of both
games \cite{Santos}.

In the present work, we study the PDG on Barab\'asi-Albert (BA) scale-free
networks \cite{Barabasi, Albert}. Serval alternative update rules determining
the evolution of each player's strategy are considered in the following. Using
systematic Monte Carlo (MC) simulations we have calculated the average density
of cooperators as a function of the temptation to defect. It is shown that the
results obtained by numerical experiments depend strongly on the dynamics of
the game, whichs suggest that the scale-topology of the underlying interacting
network may not be the crucial factor persisting the high density of the
cooperators. Of particular interesting, we have found that the system undergoes
a phase transition, from active state (coexistence of cooperators and
defectors) to absorbing state (only defectors surviving) when allowing \lq\lq
worse \rq\rq strategy (adopted by the player who gains a lower payoff) to be
imitated in the evolution of the game.

\textit{The model and simulation}. We consider the PDG with pure strategist:
the players can follow two strategies, either to defect or to cooperate ($C$).
Each player interacts with its neighbors and collects payoff depended on the
payoff parameters. The total payoff of a certain player is the sum over all
interactions. Following common studies \cite{Nowak_1, Szabo_0, Abramson, Kim,
Santos}, we also start by rescaling the game such that it depends on a single
parameter, i.e., we can choose $R=1$, $P=S=0$, and $T=b$ $(1.0 \leq b \leq
2.0)$ representing the advantage of defectors over cooperators (or the
temptation to defect), without any loss of generality of the game. After each
round, the players are allowed to inspect their neighbors' payoffs and,
according to the comparison, determine their strategies to be adopted in the
next round. To investigate how the dynamics of the game affect the evolution of
the cooperation, three kinds of update rules which determine the transformation
of each player's strategy are considered in the following.

I) \textsl{Best-takes-over}. It is commonly observed that people try to imitate
a strategy of their most successful neighbor \cite{Tomochi}. Thus we first use
a deterministic rule according to which the individual with the highest gain in
a given neighborhood reproduces with certainty. Since the PDG performed on
scale-free networks whose elements (or nodes) possess heterogeneous
connectivity \cite{Albert}. To avoid an additional bias from the higher degree
of some nodes, the gain of the certain player is calculated as the average
payoff of the individual interactions: the sum of the payoff from each neighbor
is divided by the number of the neighbors. It is important to note that this
rule does not reduce to the replicator dynamics when applied to
individual-based models of populations without spatial structure
\cite{Hauert_0}. This update rule has also been widely adopted in the studying
of PDG \cite{Zimmermann, Tomochi, Abramson, Kim}.

II) \textsl{Betters-possess-chance}. Technically, the rule I is particularly
simple to implement, but its biological relevance is rather limited because it
assumes a noise free environment. Thus in the second case the stochasticity is
add to the dynamics, and we adopt the update method just as the one used in
Ref. \cite{Hauert} (the unique place different from Ref. \cite{Santos} is to
consider average payoff of the players rather than the total payoff), i.e.,
evolution is carried out implementing the finite population analogue of
replicator dynamics \cite{Gintis, Hauert} by means of the following transition
probabilities: in each generation, whenever a site $i$ is updated, a neighbor
$j$ is drawn at random among all its neighbors; whenever $E_j>E_i$ (i.e., only
the \textit{better} players have the chance to reproduce, and if $E_i>E_j$ the
player does not change its strategy), the chosen neighbor takes over site $i$
with probability given by
\begin{equation}
W=\frac{E_j-E_i}{T-S},
\end{equation}
where $E_i$ and $E_j$ correspond to the average payoffs accumulated by the
player $i$ and $j$ in the previous round respectively.

III) \textsl{Payoff-difference-dependent}. One can see that, for both the above
rules, the \emph{error} mutation, which is very common in evolutionary systems,
is not permitted, i.e., the players who gain lower average payoffs have no
chance to replace a neighbor who does better than them. The update rule
dependent on the payoff difference, which was adopted widely in literature
\cite{Szabo_0, Szabo_1, Szabo_2, Hauert} can overcome this difficult. Given the
average payoffs ($E_i$ and $E_j$) from the previous round, the player $i$
adopts the neighbor $j$'s strategy with probability
\begin{equation}
W = \frac{1}{1 + \exp{[-(E_j - E_i)/\kappa]}},
\end{equation}
where $\kappa$ characterizes the noise introduced to permit irrational choices.
This update rule states that the strategy of a better performing player is
readily adopted, whereas it is unlikely, but not impossible, to adopt the
strategies of worse performing payers. The parameter $\kappa$ incorporates the
uncertainties in the strategy adoption. In the limit of large $\kappa$ values,
all information is lost, that is, player $i$ is unable to retrieve any
information from $E_j$ and switches to the strategy of $j$ by tossing a coin
\cite{Hauert}. Generate a random number $r$ uniformly distributed between zero
and one, if $r< W$, the neighbor's strategy is imitated.

Initially, the two strategies was randomly distributed among the players with
equal probability 1/2. The above rules of the model are iterated with parallel
updating by varying the value of $b$. The total sampling times are $11000$ MC
steps and all the results shown below are averages over the last $1000$ steps.

\begin{figure}
\centerline{\epsfxsize=8.4cm \epsffile{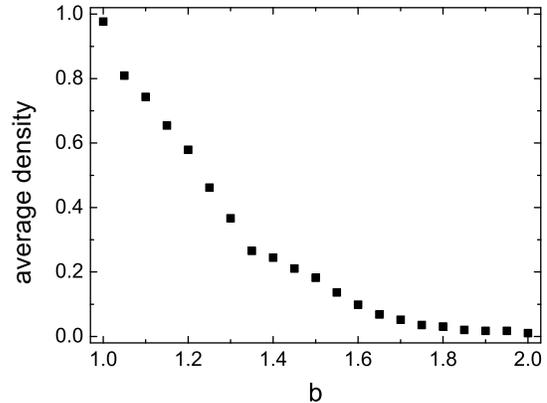}} \caption{Average density of
cooperators as a function of the temptation to defect $b$ in a evolutionary PDG
on BA scale-free networks driven by the update rule I (best-takes-over).}
\label {fig1}
\end{figure}

\begin{figure}
\centerline{\epsfxsize=8.4cm \epsffile{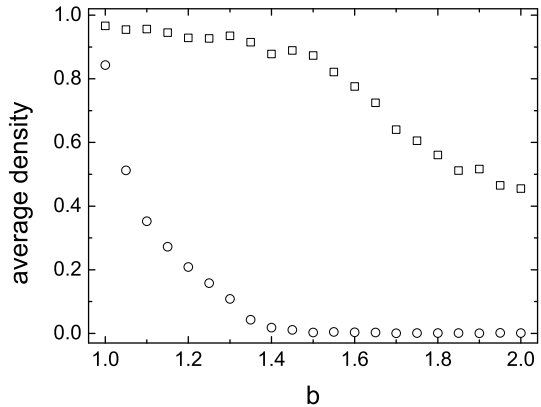}} \caption{As shown in Fig.
\ref{fig1}, but for the system driven by the update rule II
(betters-possess-chance). For the sake of comparison, the evolutionary results
adopting the original method of Ref. \cite{Santos} are also given out using
squares.} \label {fig2}
\end{figure}

\begin{figure}
\centerline{\epsfxsize=8.4cm \epsffile{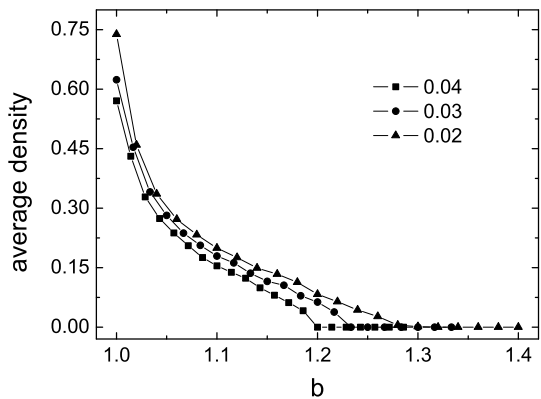}} \caption{As shown in Fig.
\ref{fig1}, but for the system driven by the update rule III
(payoff-difference-dependent). Squares, circles and triangles correspond to
different noise intensity $\kappa=0.04, 0.03, 0.02$ respectively. The lines are
guides to the eye.} \label {fig3}
\end{figure}

\begin{figure}
\centerline{\epsfxsize=8.4cm \epsffile{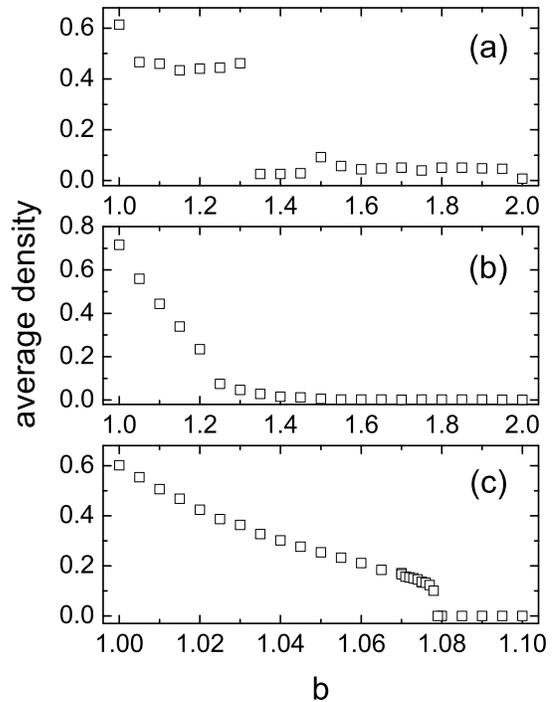}} \caption{Average density of
cooperators as a function of the temptation to defect $b$ in a evolutionary PDG
on a random regular graph with total nodes $N=10000$. The number of neighbors
is fixed as 4 for each node. From top to bottom, (a), (b) and (c) correspond to
the evolutionary results of the game driven by the update rule I, II and III
respectively. The data in (c) are obtained for $\kappa = 0.03$. } \label {fig4}
\end{figure}

\textit{Results and discussion}. In the following we show the results of
simulations performed in a system of $N=10000$ players located on BA scale-free
networks with average connectivity of the nodes fixed as 4 (the construction of
BA network can refer to Refs. \cite{Barabasi, Albert}). Our key quantity is the
average density of players adopting the strategy $C$ in the steady-state. First
let us consider the model driven by the rule I. The simulation results are
shown in Fig. \ref{fig1}. The cooperators and defectors coexist and coevolution
throughout the entire range of parameter $b$. With the increasing of the
temptation to defect, the average density of cooperators decreases
monotonically; the cooperation is inhibited quickly and sustains a low level in
a wide range of the parameter, which is clearly different from the results
obtained in Ref. \cite{Santos} using different dynamics where the cooperators
dominate the whole range of parameter of the game.

The evolutionary results of the game under the update rule II are reported in
Fig. \ref{fig2}. As compared to the former case, cooperators can exist and
survive in the whole region of $b$. However, the cooperation in the region of
large values of $b$, namely $b>1.4$, is extremely inhibited when allowing more
\lq\lq better \rq\rq players' strategies to be imitated in strategy updating of
the players. The density of cooperators maintains a minor level and is almost
invisible in Figure 2 (yet larger than zero). To compare distinctly with the
results presented in Ref. \cite{Santos}, we also calculated the average density
of $C$ by taking account into total payoff difference, just as what has been
done in Ref. \cite{Santos}, instead of average payoff difference in the update
rule II. As expected, we recover qualitatively the results of Ref.
\cite{Santos}: cooperation becomes the dominating trait throughout the entire
range of parameter of the game. The minor difference comes from the average
times of the results. Due to the computational resource limit, here the
experiment results are averaged over 20 simulations for the same network of
contacts (less than the 100 simulations in Ref. \cite{Santos}). The difference
between the two results is distinct: the cooperation is no longer dominating
whenever the average payoff difference is considered in rule II.

Now let us consider the case of payoff-difference-dependent. Figure 3 shows the
$b$ dependence of the average density of cooperators for different intensity of
the noise $\kappa=0.02, 0.03, 0.04$. Once again, the cooperation is not the
favorable choice of the players in a wide range of $b$. Of particular
interesting, one can observe that there arises two separate phases (coexistence
phase and absorbing phase) of the evolution. As indicated, the average density
of $C$ decreases monotonically with increasing $b$ until a certain threshold
value $b_c$ where the cooperators vanish and an absorbing state (all defectors)
forming. The threshold value depends on the level of the noise: the smaller
intensity of the noise $\kappa$, the larger threshold $b_c$. These phenomena
are reminiscent of the studies in Ref. \cite{Szabo_0}, where the players are
located on a two dimensional square lattice with periodic boundary and interact
with their four nearest neighbors.

From the above simulations, one can see that the dynamics (or at least the
combination of the structure of the network and the dynamics), rather than only
the scale-free topology, has an deterministic influence on the evolution of the
PDG. In order to check this statement, we also investigated the case that all
players are located on a random regular graph \cite{bollobas}. These two
network models hold similar geometric characters: small average path length and
small clustering coefficient, whereas expect for the connectivity distribution
of the vertices (one behaves scale-free and the other displays peak
distribution) \cite{Albert}. In this way, we expect that the experimental
results may give out some intrinsic view: whether the scale-free topology is
crucial for the emergence of high density of cooperators? The simulation
results obtained for random regular graph are plotted in Fig. \ref{fig4}. The
qualitative behavior of the data are consistent, on the whole, with those that
the PDG is played on BA scale-free networks. For best-takes-over rule, the lack
of stochasticity magnifies the importance of certain local configurations,
which results in discontinuous jumps and plateaus of the steady state density
of cooperators as a function of the temptation to defect $b$
(Fig.\ref{fig4}(a)), different from the continuous decay on BA scale-free
networks (Fig. \ref{fig1}). In spite of this quantitative discrepancy, however,
the qualitative trend is similar, i.e., high concentration of cooperators for
weak temptation to defect and minor level for the strong case. Whenever all
better players are allowed to share a chance to reproduce, the cooperation is
inhibited in the same way with the increasing of temptation and the average
density of $C$ maintain a very low level for large $b$ values (larger than zero
though invisible in Fig.\ref{fig4}(b)); and if bad imitation is permitted (the
rule III), absorbing phase arising when a certain value of $b_c$ arrived
(Fig.\ref{fig4}(c)).

\emph{Conclusions.} To sum up, we have explored the general question of
cooperation formation and sustainment on BA scale-free networks based on the
framework of PDG with different driving dynamics. The simulation results
suggest that the topology of the underlying interacting network, i.e., the
scale-free structure, may not be the crucial factor for the emergence and the
persistence of the cooperation, whose evolutionary results depend strongly on
the dynamics governing the game. These results are different from those
obtained in a recent work Ref. \cite{Santos} whose researches support that the
scale-free networks are advantageous for the emergence of cooperation. Of
particular interesting, we have found that the system undergoes a phase
transition, from active state to absorbing state when allowing \lq\lq worse
\rq\rq strategy to be imitated in the evolution of the game. Comparisons
between the evolution implemented on BA scale-free networks and on random
regular graph also gave out the same hints: there is no obvious evidence
supporting the scale-free topology possessing particular advantage for the
emergence of cooperation.

A lots of things are waited to do further. Here we only considered the case of
PDG. Are the results obtained in the present work also suitable for the case of
the snowdrift game when considering different dynamics? how the fraction of
cooperators goes to zero when taking account into the
payoff-difference-dependent rule? What is the accurate diagrams between $b_c$
and $\kappa$? What is the relationship between the extinction behavior of
cooperators and the case studied by Szab\'o and T\H{o}ke \cite{Szabo_0} (there
they found that the extinction behavior of the cooperators on square lattice
when increasing $b$ belongs to universality class of directed percolation)?
Work along these lines is in progress.

This work was supported by the Doctoral Research Foundation awarded by Lanzhou
University.


\begin{references}
\bibitem{Maynard}
J. Maynard Smith, E. Szathm\'ary, \emph{The Major Transitions in Evolution}
(Oxford, 1995).

\bibitem{Dugatkin}
L.A. Dugatkin, \emph{Cooperation Among Animals: An Evolutionary perspective}
(Oxford Univ. Press, Princetion, NJ, 1997).

\bibitem{Hofbauer}
J. Hofbauer and K. Sigmund, \emph{Evolutionary Games and Population
Dynamics}(Cambridge University Press, Cambridge, 1998).

\bibitem{Gintis}
H. Gintis, \emph{Game Theory Evolving} (Princeton University, Princeton, NJ,
2000).

\bibitem{Nowak}
M.A. Nowak and K. Sigmund, Nature \textbf{355}, 250 (1992)

\bibitem{Nowak_1}
M.A. Nowak and R. M. May, Nature \textbf{359}, 826 (1992); Int. J. Bifurcation
Chaos \textbf{3}, 35 (1993).

\bibitem{Szabo_1}
G. Szab\'o and C. Hauert, Phys. Rev. Lett. \textbf{89}, 118101 (2002).

\bibitem{Santos}
F.C. Santos and J.M. Pacheco, Phys. Rev. lett. \textbf{95}, 098104 (2005).

\bibitem{Zimmermann}
M.G. Zimmermann \emph{et al.}, Phys. Rev. E \textbf{69}, 065102 (2004).

\bibitem{Tomochi}
M. Tomochi and M. Kono, Phys. Rev. E \textbf{65}, 026112 (2002).

\bibitem{Barabasi}
A.-L. Barab\'{a}si and R. Albert, Science \textbf{286}, 509 (1999).

\bibitem{Albert}
R. Albert and A.-L. Barab\'{a}si, Rev. Mod. Phys. \textbf{74}, 47 (2002).

\bibitem {Abramson}
G. Abramson and M. Kuperman, Phys. Rev. E \textbf{63}, 030901 (2001).

\bibitem{Kim}
B.J. Kim \emph{et al.}, Phys. Rev. E \textbf{66}, 021907 (2002).

\bibitem{Szabo_0}
G. Szab\'o and C. T\"{o}ke, Phys. Rev. E \textbf{58}, 69 (1998).

\bibitem{Hauert_0}
C. Hauert and G. Szab\'o, Am. J. Phys. \textbf{73}, 405 (2005).

\bibitem{Hauert}
C. Hauert and M. Doebeli, Nature \textbf{428}, 643 (2004).

\bibitem{Szabo_2}
G. Szab\'o and C. Hauert, Phys. Rev. E \textbf{66}, 062903 (2002); G. Szab\'o
and J. Vukov, \emph{ibid.} \textbf{69}, 036107 (2004); J. Vukov and G. Szab\'o,
\emph{ibid.} \textbf{71}, 036133 (2005).

\bibitem{bollobas}
B. Bollob\'as, \textit{Random Graphs}, (Academic Press, New York, 1995).
\end{references}
\end{document}